\documentclass[aps,pra,preprint,showpacs,amsmath,amssymb,12pt]{revtex4-1}
\usepackage{graphicx}
\usepackage{bm}
\usepackage{amsmath}
\usepackage{slashed}
\usepackage{amssymb}

\newcommand{\iu}{{\rm i}}

\makeatletter

\newcommand{\smash@bar}[4]{%
  \smash{\rlap{\raisebox{-#3\fontdimen5#10}{$\m@th#2\mkern#4mu\mathchar'26$}}}%
}

\begin{document}

\setlength{\baselineskip}{0.7cm}

\title{Shape effects in nonlinear Thomson and Compton processes:\\ the quest for MeV high-order harmonics}

\author{M Twardy$^1$, K Krajewska$^2$ and J Z Kami\'nski$^2$}

\affiliation{$^1$ Faculty of Electrical Engineering, Warsaw University of Technology,
Pl. Politechniki 1, 00-661 Warszawa, Poland.}
\affiliation{$^2$ Institute of Theoretical Physics, Faculty of Physics, University of Warsaw,
Ho\.za 69, 00-681 Warszawa, Poland.}

\date{\today}


\bibliographystyle{iopart-num}

\begin{abstract}
Spectra of Thomson and Compton radiation, emitted during electron scattering off an
intense laser beam, are calculated using the frameworks of classical and strong-field quantum
electrodynamics, respectively. Both approaches use a plane-wave-fronted pulse approximation
regarding the driving laser beam. Within this approximation, a very good agreement between
Thomson and Compton frequency distributions is observed provided that frequencies of the emitted radiation is relatively low. 
The dependence of frequency spectra on the laser pulse envelope is analyzed. This
becomes important in the context of ultra-short pulse generation, as illustrated by 
numerical examples. \\

\noindent{\texttt{The lecture presented during the 22nd International
Laser Physics Workshop, Prague, July 15-19, 2013.}}
\end{abstract}

\maketitle

\section{Introduction}
\label{intro}

Compton scattering occurs when an electron scattered against a laser beam emits electromagnetic 
radiation~\cite{Ehlotzky,DiPiazza}. A complete description of this phenomenon is given within the framework of strong-field
quantum electrodynamics (QED) using the Furry picture~\cite{Furry}. The classical counterpart of the Compton process is called 
Thomson scattering~\cite{Umstadter1,Umstadter2}. In the Thomson approach the emitted radiation spectrum is calculated from the Lorentz-Maxwell equations with the use of the Li\'{e}nard-Wiechert 
potentials~\cite{Jackson1998,Landau}. In this paper the incident laser beam will be modeled as a plane-wave-fronted 
pulse~\cite{Neville} and both the Compton and Thomson approaches will be studied.

In many works devoted to nonlinear Compton and Thomson scattering the driving laser beam is treated as a monochromatic plane 
wave field~\cite{brown1964,goldman1964,nikishov1964,Sengupta1949,vaschaspati:A,vaschaspati:B,sarachik1970,esarey1993,Ride1995,%
Salamin1996,Salamin1997,Salamin1998,Goreslavskii1999,Panek,Ivanov,Hartin2011,Popa1,Popa2}. In fact, only a few works 
on Compton scattering, which go beyond this approximation, can be found in literature. This includes the case when the slowly-varying envelope
approximation~\cite{Narozhny} (see, also Refs.~\cite{Roshchupkin1,Roshchupkin2}) and, more recently, the plane-wave-fronted
pulse approximation~\cite{Boca2009,Mackenroth2010,Seipt,Mackenroth2011,Boca2011,Compton,Boca2012,KrajewskaNew} is used with regard to the driving laser field.
The latter is applicable when highly energetic electrons move in a laser pulse, as the action of the ponderomotive force pushing these electrons aside with respect 
to the pulse propagation direction is negligible~\cite{bulanov2011}. In this case it is assumed that the laser pulse has infinite extension in the transverse direction.
In the classical limit, on the other hand, a more accurate description of the scattering process is available. This indicates the importance of studies which 
underline the relation between quantum and classical approaches. These are of particular interest in light of various applications of Compton and
Thomson scattering, e.g., the production of ultra-short laser pulses in the x-ray domain~\cite{esarey1993}, determining the carrier envelope phase 
of intense ultra-short pulses~\cite{Mackenroth2010}, measuring the electron beam parameters~\cite{Leemans1996}, and generating coherent comb 
structures in strong-field QED for radiation and matter waves~\cite{KrajewskaNewArchive}.

Note that a comparison of Compton and Thomson scattering, 
based on a plane-wave-fronted pulse approximation, was carried out in Refs.~\cite{Mackenroth2010,Seipt,Mackenroth2011,Boca2011}. In this context,
we compare the respective spectra for pulse envelopes which consists of subpulses. We investigate the possibility of generating
coherent frequency combs. Specifically, we look at the sensitivity of these structures to a time delay between the incident subpulses. As we demonstrate, these frequency combs
can be synthetized into ultra-short pulses with a repetition rate depending on the time delay between the subpulses.

In this paper we use units such that $\hbar=1$. Numerical results are given in
relativistic units where also $m_{\rm e}=c=1$ (here, $m_{\rm e}$ is the electron rest mass).

The paper is organized as follows. In Sec.~\ref{sec:thomson:theory} we introduce the main results for Thomson scattering based on classical electrodynamics. 
In Sec.~\ref{sec:compton:theory} we introduce the Compton scattering theory arising from strong-field QED. Sec.~\ref{sec:numerical} contains numerical 
illustrations comparing Thomson and Compton spectra. The main results are summarized in Sec.~\ref{sec:summary}.

\section{Nonlinear Thomson scattering}\label{sec:thomson:theory}

By Thomson scattering we mean the process consisting in scattering of electrons by a laser beam, described entirely within the framework 
of classical mechanics and classical electrodynamics. The two most important results relevant to our considerations are the Newton-Lorentz 
equation and the frequency-angular distribution of energy radiated by an accelerating electron. The Newton-Lorentz equation~\cite{Landau,Griff}
\begin{equation}\label{newton}
\ddot{\bm r} = \frac{e}{m_{\rm e}}\sqrt{1-{\bm \beta}^2}\,
\Bigl[\bm{\mathcal{E}}(k\cdot x)-{\bm \beta}({\bm \beta}\cdot \bm{\mathcal{E}}(k\cdot x))+c\,{\bm \beta}\times \bm{\mathcal{B}}(k\cdot x)\Bigr],
\end{equation}
describes the  acceleration of the electron moving at the reduced velocity ${\bm \beta}\equiv \dot{\bm r}/c$ when placed in the electromagnetic field
generated by a laser.
Distribution of energy radiated by the electron is given by~\cite{Jackson1998}
\begin{equation}\label{thomson:distribution}
\frac{{\rm d^{3}}E_{\rm Th}}{{\rm d}\omega_{\bm K}{\rm d^{2}}\Omega_{\bm K}}
= \alpha|\mathcal{A}_{\rm Th}|^{2},
\end{equation}
where $\alpha=e^2/(4\pi\varepsilon_0 c)$ is the fine-structure constant ($\varepsilon_0$ denotes the vacuum electric permittivity) and 
\begin{equation}\label{thomson:distribution:amplitude}
\mathcal{A}_{\rm Th} 
= \frac{1}{2\pi}\int%
\frac{\bm{n_{K}}\times[(\bm{n_{K}}-\bm{\beta})\times\dot{\bm{\beta}}]}%
{(1-\bm{\beta}\cdot\bm{n_{K}})^2}%
\exp[\iu\omega_{\bm K}(t-\bm{n_{K}}\cdot\bm{r}/c)]{\rm d}t.
\end{equation}

In order to make use of this formula we must compute a specific trajectory of the scattered electron, according to Eq.~\eqref{newton}, i.e., 
we must know the electric and magnetic fields of the laser pulse. The laser pulse is specified by a shape function that we choose as follows. 
Let us assume that the total pulse consists of $N_\mathrm{rep}$ identical subpulses that are separated by the time interval $T_\mathrm{d}$ 
and each of them lasts for $T_\mathrm{sub}$ and contains $N_\mathrm{osc}$ laser field oscillations of the frequency $\omega_\mathrm{L}$. 
This means that $\omega_\mathrm{L}T_\mathrm{sub}=2\pi N_\mathrm{osc}$ and for the envelope function we choose the sine-squared function. 
For the time interval $0\leqslant t\leqslant T_\mathrm{d}+T_\mathrm{sub}$ we define the function
\begin{equation}
F(t)=\begin{cases}
\sin^2\Bigl(\pi\frac{t-T_\mathrm{d}/2}{T_\mathrm{sub}}\Bigr)\sin(\omega_\mathrm{L}(t-T_\mathrm{d}/2)+\chi), 
& T_\mathrm{d}/2\leqslant t\leqslant T_\mathrm{d}/2+T_\mathrm{sub}, \cr
0, & \mathrm{otherwise},
\end{cases}
\label{tt1}
\end{equation}
and repeat it $N_\mathrm{rep}$ times. In this equation the real parameter, $\chi$, denotes a carrier envelope phase. Now, we introduce 
the frequency $\omega=2\pi/T_\mathrm{p}$ with $T_\mathrm{p}=N_\mathrm{rep}(T_\mathrm{d}+T_\mathrm{sub})$, and define the shape function 
$f(\phi)$ for $0\leqslant\phi\leqslant 2\pi$ such that its derivative over the phase $\phi$ equals
\begin{equation}
f'(\phi)=N_{\mathrm{A}}F(\phi/\omega).
\label{tt2}
\end{equation}

From now on, we use the Coulomb gauge for the radiation field, in which case the 
electric and magnetic field components are equal to
\begin{align}
\bm{\mathcal{E}}(k\cdot x)&=-\partial_t \bm{A}(k\cdot x)= -ck^0 \bm{A}'(k\cdot x), \label{electric} \\
\bm{\mathcal{B}}(k\cdot x)&=\bm{\nabla}\times \bm{A}(k\cdot x)= -\bm{k}\times \bm{A}'(k\cdot x). \label{magnetic} 
\end{align}
Because the electric field generated by lasers has to fulfill the following condition,
\begin{equation}
\int_{-\infty}^{\infty}\bm{\mathcal{E}}(ck^0t-\bm{k}\cdot\bm{r})\mathrm{d}t=0, \label{LaserCondition1}
\end{equation}
we have also that
\begin{equation}
\lim\limits_{t\rightarrow -\infty}\bm{A}(ck^0t-\bm{k}\cdot\bm{r})=
\lim\limits_{t\rightarrow \infty}\bm{A}(ck^0t-\bm{k}\cdot\bm{r}), \label{LaserCondition2}
\end{equation}
and, hence, we can assume that in the remote past and far future the vector potential vanishes.
Therefore, for the electromagnetic potential we choose 
\begin{equation}
A(k\cdot x)=A_0N_\mathrm{osc}\varepsilon f(k\cdot x), \label{LaserVectorPotential}
\end{equation}
with the shape function $f(k\cdot x)$ such that $f(k\cdot x)=0$ for $k\cdot x < 0$ and for $k\cdot x > 2\pi$.
In addition, $\varepsilon$ is the linear polarization vector of the laser field such that $\varepsilon^2=-1$ and $k\cdot\varepsilon=0$. Moreover, we define the dimensionless and relativistically invariant parameter
\begin{equation}
\mu=\frac{|eA_0|}{m_\mathrm{e}c},
\label{miu}
\end{equation}
which determines the intensity of the laser pulse.

Note that the shape function~\eqref{tt2} determines the electric and magnetic fields 
of the laser pulse, Eqs.~\eqref{electric} and \eqref{magnetic}. Thus, the shape function 
for the electromagnetic potential equals
\begin{equation}
f(k\cdot x)=\int_0^{k\cdot x} \mathrm{d}\phi f^{\prime}(\phi), \label{shapefunctionA}
\end{equation}
and, as desired, vanishes for $k\cdot x < 0$ and $k\cdot x > 2\pi$. In Eq.~\eqref{tt2},
the normalization constant $N_{\mathrm{A}}$ is defined such that
\begin{equation}
\frac{1}{2\pi}\int_0^{2\pi}{\rm d}\phi\,[f'(\phi)]^2=\frac{1}{2}.
\label{a3}
\end{equation}

\begin{figure}
\begin{center}
\includegraphics[width=13cm]{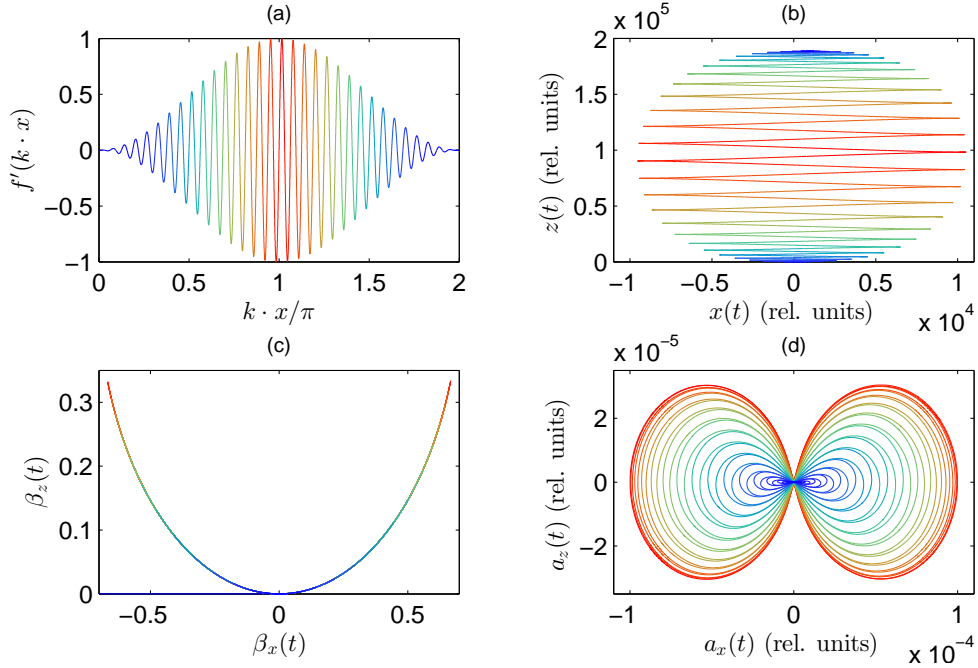}
\caption{(Color online) In panel (a), the shape function $f'(k\cdot x)$ defining the electric and magnetic field components of the 32-cycle 
laser pulse is shown for $T_\mathrm{d}=0$ and $N_\mathrm{rep}=1$. In the remaining panels, the solution of the relativistic Newton-Lorentz 
equation~(\ref{newton}) for the electron moving in the depicted laser pulse is presented (i.e., electron position, speed, and acceleration). 
The parameters of the pulse are such that $\mu=1$, $\omega_{\rm L}=10^{-4} m_{\rm e}c^2$, and $N_{\rm osc}=32$, where the scaled amplitude of vector potential,
$\mu$, measures the peak intensity of the laser field. Initially, the electron is at rest in the center of the coordinate system. \label{lorent4}}
\end{center}
\end{figure}

In our investigations we use the plane-wave-fronted pulse approximation. An applicability of this approximation 
is based on the assumption that the transverse variation of the electron trajectory in the laser field is negligible 
when compared to the size of the laser focus, as illustrated in Fig.~\ref{lorent4}. In Fig.~\ref{lorent4}(a), we 
present the solution of the relativistic Newton-Lorentz equation~(\ref{newton}) for the electron initially at rest 
and in the presence of the laser pulse. The carrier frequency of the laser field $\omega_{\rm L}$ is equal to $10^{-4}m_{\rm e}c^2$ 
(which corresponds to the laser photon energy of about 50 eV), $N_{\rm osc}=32$, $N_{\rm rep}=1$, $T_\mathrm{d}=0$, and $\mu=1$. 
The conditions are set such that initially the electron is at rest in the center of the coordinate system, so its velocity is zero 
before the arrival of the laser pulse. Fig.~\ref{lorent4}(b) depicts the $x$- and $z$-coordinates of the electron's position,  
Fig.~\ref{lorent4}(c) -- the reduced velocity, and Fig.~\ref{lorent4}(d) -- the acceleration of the electron. The colors of 
these distributions correspond to the colors of the laser pulse [Fig.~\ref{lorent4}(a)]. It turns out, that for higher amplitudes of the field depicted in
Fig.~\ref{lorent4}(a), the trajectory of the electron preserves its shape form [Fig.~\ref{lorent4}(b)]; however, 
the values of the speed components $\beta_{x}(t)$ and $\beta_{z}(t)$ [Fig.~\ref{lorent4}(c)] attain higher values.

As one can see from Fig.~\ref{lorent4}, the classical electron placed in a linearly polarized laser field
exhibits an oscillatory motion along the direction of the electric field component (i.e., along the $x$-axis)
together with a drift motion in the propagation direction of the laser pulse (i.e., in the $z$-direction). It may also be 
observed that for a given laser field frequency the displacement of the electron is larger for a stronger laser field  
than for a weaker laser field. A useful measure of the relativistic length unit is the reduced electron
Compton wavelength 
\[
\lambdabar_{\rm C} = \frac{\hbar}{m_{\rm e}c} = 3.8616\times10^{-13}\,{\rm m}\,,
\]
which equals 1 in relativistic units. Note that for the chosen laser field parameters the electron displacement 
along the electric field vector is of the order of $10^{-9}$m (Fig.~\ref{lorent4}). 
Taking this into account, we find that the electron displacement along the electric field
direction is of the order of $0.001\mu$m. For lasers available today, a typical linear dimension of their focus is a few $\mu$m. Therefore, the electron displacement
in the transverse direction (even for very powerful laser fields) can be neglected on the scale of the focus, provided that the 
laser frequency is sufficiently large in the reference frame in which initially electrons are at rest. We conclude
that the plane-wave-fronted pulse approximation is perfectly suitable for describing the nonlinear Thomson scattering processes generated by currently available laser sources.

\section{Nonlinear Compton scattering}\label{sec:compton:theory}

Using the $S$-matrix formalism of strong-field QED, we derive that the probability amplitude for 
the Compton process, $e^-_{\bm{p}_{\mathrm{i}}\lambda_{\mathrm{i}}}\rightarrow e^-_{\bm{p}_{\mathrm{f}}\lambda_{\mathrm{f}}}+\gamma_{\bm{K}\sigma}$, 
with the initial and final electron momenta and spin polarizations $\bm{p}_{\mathrm{i}}\lambda_{\mathrm{i}}$ and $\bm{p}_{\mathrm{f}}\lambda_{\mathrm{f}}$, 
respectively, equals
\begin{equation}
{\cal A}(e^-_{\bm{p}_{\mathrm{i}}\lambda_{\mathrm{i}}}\rightarrow e^-_{\bm{p}_{\mathrm{f}}\lambda_{\mathrm{f}}}
+\gamma_{\bm{K}\sigma})=-\mathrm{i}e\int \mathrm{d}^{4}{x}\, j^{(++)}_{\bm{p}_{\mathrm{f}}\lambda_{\mathrm{f}},
\bm{p}_{\mathrm{i}}\lambda_{\mathrm{i}}}(x)\cdot A^{(-)}_{\bm{K}\sigma}(x), \label{ComptonAmplitude}
\end{equation}
where $\bm{K}\sigma$ denotes the Compton photon momentum and polarization. Here, we consider the case when both the laser pulse and the Compton photon
are linearly polarized. In Eq.~\eqref{ComptonAmplitude},
\begin{equation}
A^{(-)}_{\bm{K}\sigma}(x)=\sqrt{\frac{1}{2\varepsilon_0\omega_{\bm{K}}V}} 
\,\varepsilon_{\bm{K}\sigma}\mathrm{e}^{\mathrm{i}K\cdot x},
\label{per}
\end{equation}
where $V$ is the quantization volume, 
$\omega_{\bm{K}}=cK^0=c|\bm{K}|$ ($K\cdot K=0$), and $\varepsilon_{\bm{K}\sigma}=(0,\bm{\varepsilon}_{\bm{K}\sigma})$ 
are the polarization four-vectors satisfying the conditions
\begin{equation}
K\cdot\varepsilon_{\bm{K}\sigma}=0,\quad \varepsilon_{\bm{K}\sigma}\cdot\varepsilon_{\bm{K}\sigma'}=-\delta_{\sigma\sigma'},
\end{equation}
for $\sigma,\sigma'=1,2$. Moreover, $j^{(++)}_{\bm{p}_{\mathrm{f}} \lambda_{\mathrm{f}},\bm{p}_{\mathrm{i}}\lambda_{\mathrm{i}}}(x)$ 
is the matrix element of the electron current operator with its $\nu$-component equal to
\begin{equation}
[j^{(++)}_{\bm{p}_{\mathrm{f}} \lambda_{\mathrm{f}},\bm{p}_{\mathrm{i}}\lambda_{\mathrm{i}}}(x)]^{\nu}
=\bar{\psi}^{(+)}_{\bm{p}_{\mathrm{f}} \lambda_{\mathrm{f}}}(x)\gamma^\nu \psi^{(+)}_{\bm{p}_{\mathrm{i}}\lambda_{\mathrm{i}}}(x).
\end{equation}
Here, $\psi^{(+)}_{\bm{p}\lambda}(x)$ is the so-called Volkov solution of the Dirac equation coupled to the electromagnetic field~\cite{Volkov,KK} (see, also Refs.~\cite{Varro:A,Varro:B,Raicher2013}
for possible generalizations)
\begin{equation}
\psi^{(+)}_{\bm{p}\lambda}(x)=\sqrt{\frac{m_{\mathrm{e}}c^2}{VE_{\bm{p}}}}\Bigl(1-\frac{e}{2k\cdot p}\slashed{A}\slashed{k}\Bigr)
u^{(+)}_{\bm{p}\lambda}\mathrm{e}^{-\mathrm{i} S_p^{(+)}(x)} , \label{Volk}
\end{equation}
with
\begin{equation}
S_p^{(+)}(x)=p\cdot x+\int^{k\cdot x} \Bigl[\frac{ e A(\phi )\cdot p}{k\cdot p}
-\frac{e^2A^{2}(\phi )}{2k\cdot p}\Bigr]{\rm d}\phi .
\label{bbb}
\end{equation}
Moreover, $E_{\bm{p}}=cp^0$, $p=(p^0,\bm{p})$, $p\cdot p=m_{\mathrm{e}}^2c^2$, and $u^{(+)}_{\bm{p}\lambda}$ 
is the free-electron bispinor normalized such that
\begin{equation}
\bar{u}^{(+)}_{\bm{p}\lambda}u^{(+)}_{\bm{p}\lambda'}=\delta_{\lambda\lambda'}.
\end{equation}
The four-vector potential $A(k\cdot x)$ in Eq.~\eqref{Volk} represents an external electromagnetic 
radiation generated by lasers, in the case when a transverse variation of the laser field in a focus is
negligible. In other words, $A(k\cdot x)$ represents the plane-wave-fronted pulse. In this case, $k\cdot A(k\cdot x)=0$ and $k\cdot k=0$, 
which allows one to exactly solve the Dirac equation for such electromagnetic fields. 

The probability amplitude for the Compton process~\eqref{ComptonAmplitude} becomes
\begin{equation}
{\cal A}(e^-_{\bm{p}_\mathrm{i}\lambda_\mathrm{i}}\longrightarrow e^-_{\bm{p}_\mathrm{f}\lambda_\mathrm{f}}+\gamma_{\bm{K}\sigma})
=\mathrm{i}\sqrt{\frac{2\pi\alpha c(m_{\mathrm{e}}c^2)^2}{E_{\bm{p}_\mathrm{f}}E_{\bm{p}_\mathrm{i}}\omega_{\bm{K}}V^3}}\, \mathcal{A}, \label{ct1}
\end{equation}
where 
\begin{equation}
\mathcal{A}= \int\mathrm{d}^{4}{x}\,   \bar{u}^{(+)}_{\bm{p}_\mathrm{f}\lambda_\mathrm{f}}\Bigl(1-\mu\frac{m_\mathrm{e}c}{2p_\mathrm{f}\cdot k}f(k\cdot x)\slashed{\varepsilon}\slashed{k}\Bigr)\slashed{\varepsilon}_{\bm{K}\sigma}  \,
\Bigl(1+\mu\frac{m_\mathrm{e}c}{2p_\mathrm{i}\cdot k} f(k\cdot x)\slashed{\varepsilon}\slashed{k}\Bigr) u^{(+)}_{\bm{p}_\mathrm{i}\lambda_\mathrm{i}}\,\mathrm{e}^{-\mathrm{i}S(x)},  \label{ct2}
\end{equation}
with
\begin{equation}
S(x) = S^{(+)}_{{p}_\mathrm{i}}(x)-S^{(+)}_{{p}_\mathrm{f}}(x)-K\cdot x\,.
\end{equation}

While moving in a laser pulse, the electron acquires an additional momentum shift~\cite{Compton,BW}, this leads  to a notion of the laser-dressed momentum:
\begin{equation}
\bar{p} = p-\mu m_{\rm e}c\frac{p\cdot\varepsilon}{p\cdot k}\left\langle f\right\rangle k 
+\frac{1}{2}(\mu m_{\rm e}c)^2\frac{1}{p\cdot k}\left\langle f^2\right\rangle k\,.
\end{equation}
Having this in mind we can define
\begin{equation}
N_{\rm eff} = \frac{K^{0}+\bar{p}^{0}_{\rm f}-\bar{p}^{0}_{\rm i}}{k^{0}}=
cT_{\rm p}\frac{K^{0}+\bar{p}^{0}_{\rm f}-\bar{p}^{0}_{\rm i}}{2\pi},
\end{equation}
which is both gauge- and relativistically invariant~\cite{Compton}.

The frequency-angular distribution of energy of the emitted photons for an unpolarized electron beam is given by
\begin{equation}\label{copton:spectrum:3}
\frac{{\rm d^{3}}E_{\rm C}}{{\rm d}\omega_{\bm K}{\rm d^{2}}\Omega_{\bm K}}
=\frac{1}{2}\sum_{\sigma=1,2}\sum_{\lambda_{\rm i}=\pm}\sum_{\lambda_{\rm f}=\pm}
\frac{{\rm d^{3}}E_{{\rm C},\sigma}(\lambda_{\rm i},\lambda_{\rm f})}{{\rm d}\omega_{\bm K}{\rm d^{2}}\Omega_{\bm K}},
\end{equation}
where
\begin{equation}\label{copton:spectrum:2}
\frac{{\rm d^{3}}E_{{\rm C},\sigma}(\lambda_{\rm i},\lambda_{\rm f})}{{\rm d}\omega_{\bm K}{\rm d^{2}}\Omega_{\bm K}}
=\alpha\left|\mathcal{A}_{{\rm C},\sigma}(\omega_{\bm{K}},\lambda_{\rm i},\lambda_{\rm f})\right|^2,
\end{equation}
and the scattering amplitude equals
\begin{equation}\label{copton:spectrum:1}
\mathcal{A}_{{\rm C},\sigma}(\omega_{\bm{K}},\lambda_{\rm i},\lambda_{\rm f}) 
=\frac{m_{\rm e}c K^{0}}{2\pi\sqrt{p_{\rm i}^{0}k^{0}(k\cdot p_{\rm f})}}
\sum_{N}D_{N}\frac{1-{\rm e}^{-2\pi\iu(N-N_{\rm eff})}}{\iu(N-N_{\rm eff})},
\end{equation}
with the functions $D_N$ defined in~\cite{Compton}.

\begin{figure}
\begin{center}
\includegraphics[width=11cm,clip]{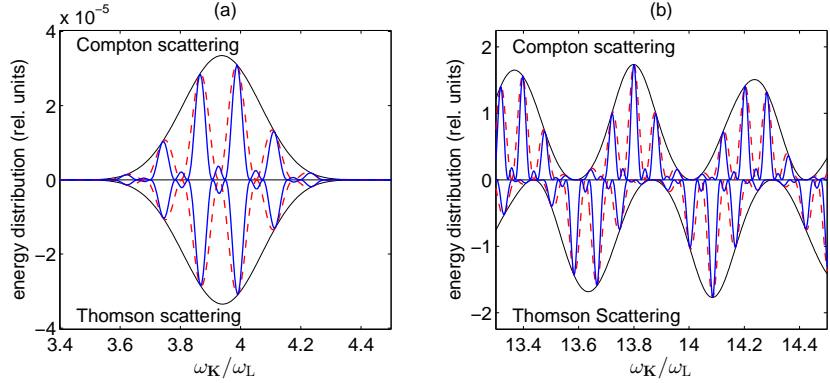}
\caption{\label{p1:tcomb220130518fMirrorCom} (Color online) The energy distribution of radiation emitted by the scattered electron. 
The black solid line corresponds to $N_{\text{rep}}=1$, the red dashed line to $N_{\text{rep}}=2$, whereas the blue solid line is 
for $N_{\text{rep}}=3$. Each subpulse contains eight field oscillations ($N_{\rm osc}=8$). The detection angles are $\varphi_{\bm K}=0$ and $\theta_{\bm K}=\pi/10$.
While the results presented in panel (a) relate to $\mu = 1$ and $\omega_{\text{L}} = 3\times 10^{-3}m_{\text{e}}c^2$, the results presented in
panel (b) are for $\mu = 10$ and $\omega_{\text{L}} = 3\times 10^{-2}m_{\text{e}}c^2$.
}
\end{center}
\end{figure}

In Fig.~\ref{p1:tcomb220130518fMirrorCom}, we compare the Thomson and Compton energy distributions for a single laser pulse 
in the reference frame in which initially electrons are at rest. As expected, for $\omega_{\bm{K}}\ll m_\mathrm{e}c^2$ 
both theories give the same results, as depicted in the left panel. However, for larger laser carrier frequency $\omega_\mathrm{L}$ 
and energies of generated photons $\omega_{\bm{K}}$ the Compton distribution is red-shifted with respect to the Thomson one. 
Such a shift has been analyzed in~\cite{Seipt} within the slowly changing envelope approximation. This analysis has been 
extended to arbitrarily short laser pulses in~\cite{KrajewskaNewArchive2}, together with the discussion of the significant 
role played by the polarization of emitted radiation and the spin degrees of freedom of the electron initial and final states.

\begin{figure}
\begin{center}
\includegraphics[width=11cm]{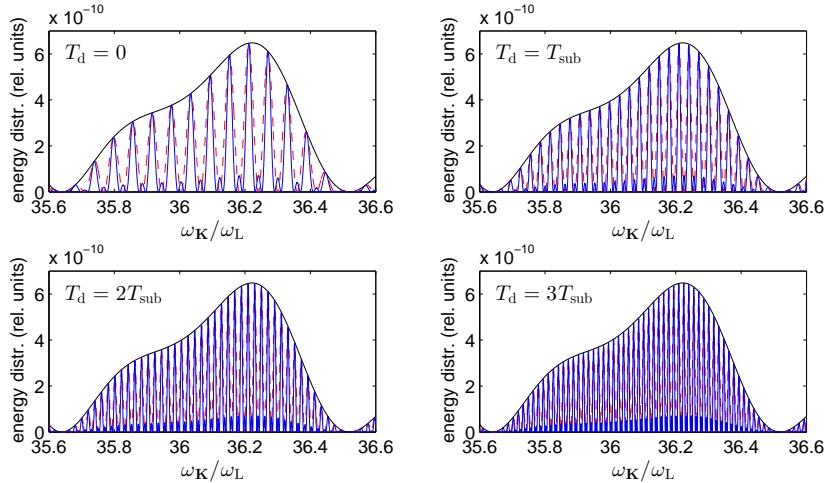}
\caption{\label{combdall_20131009ex}  (Color online) The Compton energy distribution for the laser pulse parameters: 
$\omega_{\rm L} = 4.15\times 10^{-4}m_{\rm e}c^2$, $\mu = 1$, $N_{\rm osc} = 16$, $\theta_{\bm{K}}=0.2\pi$, $\varphi_{\bm{K}}=0$, 
and $\chi=\pi$. Different panels correspond to four different delay times $T_\mathrm{d}$. The solid black lines (envelopes) 
present the energy distributions for a single pulse ($N_\mathrm{rep}=1$), the dashed red lines are for $N_\mathrm{rep}=2$, whereas the solid 
blue lines are for $N_\mathrm{rep}=3$. The corresponding energy distributions are divided by $N_\mathrm{rep}^2$, which proves the coherent properties of the generated high-order harmonics comb structures.
}
\end{center}
\end{figure}

\section{Compton high-order harmonics and generation of ultra-short pulses}
\label{sec:numerical}

High-order harmonics generated via non-relativistic interaction of intense laser pulses with atoms allow to synthesize attosecond 
pulses of coherent radiation~\cite{Farkas}. Currently, the energy bandwidth of the harmonic plateau can reach a few keV~\cite{Popmintchev}. In order to extend this 
spectrum up to MeV domain a relativistic treatment is necessary. The Compton process offers such a possibility, as it is presented 
in Fig.~\ref{combdall_20131009ex}. For a single laser pulse ($N_{\rm rep}=1$), we observe a broad and smooth energy distribution from which we choose 
a part of the bandwidth approximately equal to the carrier frequency $\omega_\mathrm{L}$. However, if we apply the sequence of 
$N_\mathrm{rep}$ pulses the energy distribution shows an equally spaced peaks with maxima which scale as $N_{\mathrm{rep}}^2$. 
This scaling law indicates that the generated comb structure is temporarily coherent. Moreover, the distance between the peaks 
can be controlled by a delay of the laser subpulses. Therefore, one can interpret the emergence of such a structure as the result 
of the interference of Compton photons emitted from different subpulses~\cite{KrajewskaNewArchive}.

\begin{figure}
\begin{center}
\includegraphics[width=12cm]{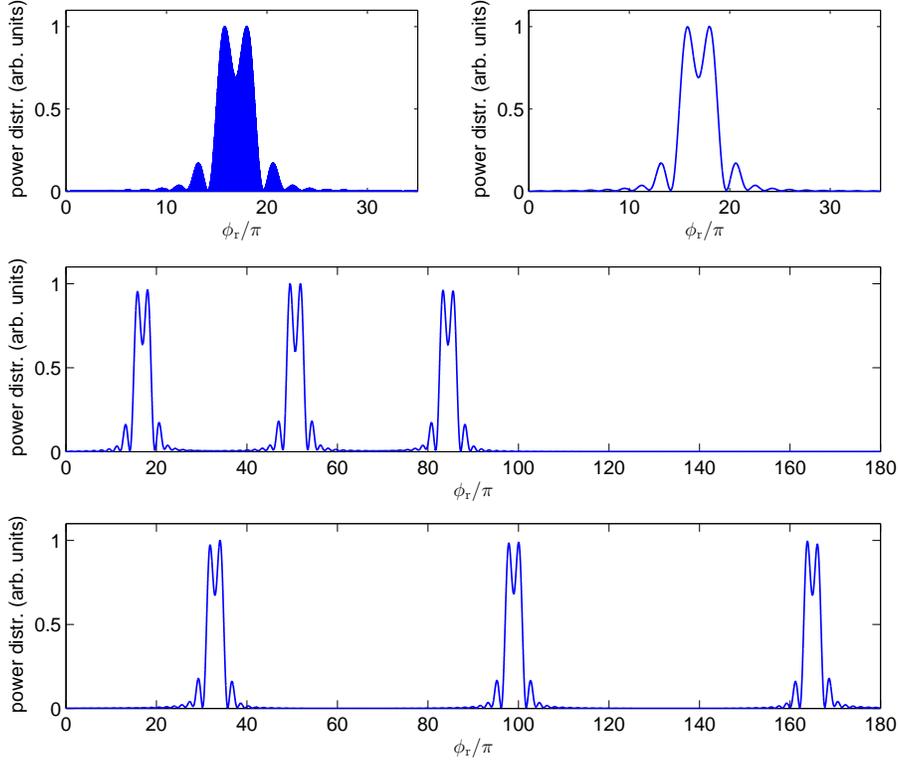}
\caption{\label{qpuls16tc3} (Color online) Temporal power distributions of generated radiation, synthesized from the Compton amplitude~\eqref{copton:spectrum:1} 
for $\lambda_\mathrm{i}\lambda_\mathrm{f}=1$ and for the laser field parameters specified in Fig.~\ref{combdall_20131009ex}. The upper two panels show the 
power distributions defined by Eqs.~\eqref{tt9} and \eqref{ttt8} for a single pulse ($N_{\rm rep}=1$), after being normalized to the maximum value. 
The middle and the bottom panels show the temporal power distributions~\eqref{ttt8} composed from the energy distributions represented in Fig.~\ref{combdall_20131009ex} by the blue lines ($N_\mathrm{rep}=3$) for $T_\mathrm{d}=0$ and $T_\mathrm{d}=T_\mathrm{sub}$, respectively.
}
\end{center}
\end{figure}

In order to further investigate the coherent properties of such a high-order harmonic spectrum let us consider 
the temporal power distribution of emitted radiation. This power distribution is related to the Compton 
amplitude~\eqref{copton:spectrum:1} by the formula
\begin{equation}
\frac{\mathrm{d}^2P_{\mathrm{C},\sigma}(\phi_{\mathrm{r}},\lambda_{\rm i},\lambda_{\rm f})}{\mathrm{d}^2\Omega_{\bm{K}}}=\frac{\alpha}{\pi}\bigl(\Re \tilde{\mathcal{A}}^{(+)}_{\mathrm{C},\sigma}(\phi_{\mathrm{r}},\lambda_{\rm i},\lambda_{\rm f})\bigr)^2.
\label{tt9}
\end{equation}
where
\begin{equation}
\tilde{\mathcal{A}}^{(+)}_{\mathrm{C},\sigma}(\phi_{\mathrm{r}},\lambda_{\rm i},\lambda_{\rm f})=\int_0^{\infty}\mathrm{d}\omega \mathcal{A}_{\mathrm{C},\sigma}(\omega,\lambda_{\rm i},\lambda_{\rm f})\mathrm{e}^{-\mathrm{i}\omega\phi_{\mathrm{r}}/\omega_\mathrm{L} }.
\label{tt6}
\end{equation}
Here, $\Re$ denotes the real part and $\phi_{\mathrm{r}}=\omega_\mathrm{L}(t-R/c)$, with $R$ being a distance from the scattering region to the observation point. 
In general, the power distribution~\eqref{tt9} is a very rapidly oscillating function of time. One can define the temporal power distribution of generated radiation, 
avaraged over these oscillations, 
\begin{equation}
\frac{\mathrm{d}^2\langle P_{\mathrm{C},\sigma}\rangle(\phi_{\mathrm{r}},\lambda_{\rm i},\lambda_{\rm f})}{\mathrm{d}^2\Omega_{\bm{K}}}=\frac{\alpha}{2\pi} |\tilde{\mathcal{A}}^{(+)}_{\mathrm{C},\sigma}(\phi_{\mathrm{r}},\lambda_{\rm i},\lambda_{\rm f})|^2 .
\label{ttt8}
\end{equation}

Fig.~\ref{qpuls16tc3} depicts the temporal power distributions,~\eqref{tt9} or \eqref{ttt8}, as functions of the dimensionless retarded phase $\phi_{\mathrm{r}}=\omega_\mathrm{L}(t-R/c)$, instead of the observation time $t$. The power distributions have been synthesized from the energy distributions presented in Fig.~\ref{combdall_20131009ex}. As we see, the radiation is emitted in the form of short pulses. Similar conclusions can be drawn from the classical Thomson scattering, although some significant discrepancies between the classical and quantum theories can be observed~\cite{KTK}. 

\section{Conclusions}\label{sec:summary}

In this paper, we have presented the theory of Thomson and Compton processes in intense laser pulses. 
We have shown that, by applying the sequence of laser pulses, it is possible to create high-order harmonics 
structures in the emitted radiation. We have investigated this problem in the electron beam reference frame. 
It appears, however, that in the laboratory frame, when electrons have the energy of the order of GeV, 
the bandwidth of the generated harmonic spectrum is of the order of a few MeV. This can be further synthesized 
into zepto- or even yoctosecond pulses, as will be presented elsewhere~\cite{KTK}.

\section*{Acknowledgements}
This work is supported by the Polish National Science Center (NCN) under Grant No. 2012/05/B/ST2/02547.

\bibliography{refer}

\end{document}